\documentclass[aps, prb, twocolumn, reprint, amsmath, amssymb]{revtex4-2}
\usepackage{graphicx}
\usepackage{hyperref}
\usepackage{bm}
\usepackage{subcaption}
\usepackage{dcolumn}
\usepackage{tikz}
\usetikzlibrary{patterns, arrows.meta}

\usepackage{caption}
\usepackage[margin=2.5cm]{geometry}

\hypersetup{colorlinks=true, linkcolor=blue, citecolor=blue, urlcolor=blue}

\begin{document}
	
	\title{Superconducting Diode Effect in Josephson $\varphi_0$ Junction}
	
	\author{A. Janalizadeh$^{1}$}
	\author{Y. M. Shukrinov$^{2,3}$}
	\author{M. R. Kolahchi$^{1}$}    
	\affiliation{$^{1}$Department of Physics, Institute for Advanced Studies in Basic Sciences (IASBS), Zanjan, Iran\\
		$^{2}$BLTP, JINR, Dubna, Moscow Region, 141980, Russia\\
		$^{3}$Dubna State University, Dubna, 141980, Russia}
		\date{September 11, 2026}

	\begin{abstract}
		We investigate the superconducting diode effect in a Josephson junction with a ferromagnetic weak link in the presence of Rashba spin-orbit coupling. While the standard $\varphi_0$ junction model is valid only for weak exchange fields $h \lesssim T_c$, realistic ferromagnetic barriers typically operate in the regime $h \gg T_c$, where the conventional Ginzburg-Landau gradient expansion becomes insufficient and requires the inclusion of higher order gradient terms to ensure stability of the free energy. By extending the free energy with such higher order term, we demonstrate that the linear gradient coupling alone leads to finite momentum Cooper pairing while leaving the critical current symmetry unchanged. The interplay between this linear term and the higher order gradient one leads to the Josephson diode effect.
	\end{abstract}

	\maketitle
	\newpage
\section{Introduction}
The superconducting diode effect (SDE) manifests as a directional dependence of the critical current, \(I_{c+} \neq |I_{c-}|\), such that the junction remains superconducting for one current direction and switches to a dissipative ohmic state for the opposite direction \cite{Ando2020,Nadeem2023,Nadeem2025}. This nonreciprocal response requires simultaneous breaking of inversion (\(\mathcal{P}\)) and time reversal (\(\mathcal{T}\)) symmetries. In noncentrosymmetric superconductors, the Ginzburg-Landau (GL) free energy may include a linear in the order parameter gradient, which couples the phase to an inversion breaking field such as spin-orbit coupling (SOC) \cite{MineevSamokhin2008,Mineev2008,edelstein1996,LifshitzInvariant2016}. This term shifts the free energy minimum to a finite Cooper pair momentum \(q_0\) but does not, by itself, break the symmetry between critical currents in opposite directions.

The SDE attracts much attention today and it has been observed in different Josephson structures, including topological semimetals \cite{Pal2022}, high-mobility InSb nanoflags \cite{Turini2022}, between Bi\(_2\)Sr\(_2\)CaCu\(_2\)O\(_{8+x}\) flakes twisted by 45\(^{\circ}\) across the superconducting dome \cite{Zhu2023}, and in three-terminal Josephson devices \cite{Gupta2023}. Also, high-temperature Josephson diode was investigated in Ref.\,\cite{Ghosh2024} and in symmetric Josephson junctions in Ref.\,\cite{Baumgartner2022}. Josephson diode effect based on breaking of time-reversal and inversion symmetries was demonstrated in type-II Dirac semimetals and systems with van der Waals Josephson barrier \cite{Sivakumar2024,Kim2024}. Recently, series of works were devoted to field-free Josephson diode based on asymmetric geometries, ferromagnetic barriers and exotic materials \cite{Golod2022,Chen2024,Hou2023,Guarcello2024,Jeon2022,Wu2022,Trahms2023,DiezMerida2023,Ma2025}.

The Josephson diode effect is not limited to the asymmetry of the critical currents. Other features of the current-voltage characteristics can also exhibit nonreciprocal behavior, in particular, asymmetries in the retrapping currents \cite{Seleznev2024,Misaki2021,Steiner2023,Wu2022} and in Shapiro steps \cite{Li2024,Ciaccia2024,Leblanc2024,Valentini2024} have been reported. Investigating these additional signatures can provide further insights into Josephson junction systems and their potential applications.

Buzdin proposed the \(\varphi_0\) junction, in which the weak link is a magnetic metal with broken inversion symmetry, such as MnSi or FeGe, where the linear gradient coupling yields a current-phase relation \(I(\varphi) = I_c \sin(\varphi + \varphi_0)\) \cite{Buzdin2008}. Here \(\varphi_0\) is proportional to the exchange field \(h\) and the SOC strength. This description is valid for small \(h\) within the conventional gradient expansion. Recent experiments on HgTe-based Josephson junctions have observed large anomalous phase shifts, while gate controlled InAs/Al interferometers have directly linked the phase shift to diode efficiency \cite{Huttner2026,Reinhardt2024}.

Several theoretical mechanisms for the SDE have been proposed. Kochan \(et. al.\) studied quasi two dimensional superconductors with a linear gradient coupling and an external in plane magnetic field \cite{Kochan2023}. He \(et. al.\) extended GL theory with Rashba SOC and a Zeeman field \cite{He2022}. Yuan and Fu proposed an intrinsic mechanism based on finite momentum pairing in noncentrosymmetric materials \cite{PNAS}. Meyer and Houzet analyzed a ballistic Rashba nanowire \cite{Meyer2024}, and Ili\'{c} \(et. al.\) explored diffusive Rashba structures near the \(0\)-\(\pi\) transition \cite{Ilic2024}. R\"{o}sch \(et al.\) demonstrated that magnetization gradients in Rashba metals can stabilize a helical state and enable SDE even without SOC \cite{Rosch2024}. These approaches typically rely on external fields or specific band structure details.   

Experimentally, the SDE has been observed in diverse platforms, including InAs/Al heterostructures \cite{Lotfizadeh2024}, few-layer MoTe\(_2\) \cite{Wakamura2024}, HgTe quantum wells \cite{Huttner2026}, van der Waals Josephson junctions with \(T_d\)-WTe\(_2\) \cite{Kim2024}, and zero field Fe\(_3\)GeTe\(_2\)/NbSe\(_2\) heterostructures \cite{Hu2024}.

In ferromagnets, the exchange field often exceeds \(T_c\) by orders of magnitude, necessitating a generalization of the GL expansion to include higher order gradient terms, as in the theory of Fulde-Ferrell-Larkin-Ovchinnikov (FFLO) states \cite{Buzdin2005,FFLO,Mandal2024}. In the present work, we consider the full nonlinear structure of the Ginzburg-Landau free energy without invoking the simplifications used in Ref.\cite{Buzdin2008}. Here we combine Buzdin's \(\varphi_0\) junction model with a higher order gradient term to describe the SDE beyond the small \(h\) limit. We derive the nonreciprocity, and obtain the diode quality factor \(Q\). The obtained results establish the connection between the \(\varphi_0\) junction concept and recent SDE theories \cite{He2022,Kochan2023}.

	\section{Ginzburg-Landau free energy with Rashba type spin-orbit coupling}
	
	We start from the Ginzburg-Landau (GL) free energy density for a superconductor in contact with a ferromagnetic metal with Rashba spin-orbit coupling \cite{Buzdin2008,MineevSamokhin2008}:
		\begin{equation}
		\begin{aligned}
			F_{\varphi_0} ={}& a|\psi|^2 + \gamma|\hat{D}\psi|^2 + \frac{b}{2}|\psi|^4 \\
& - \epsilon\,\mathbf{n}\!\cdot\!\left\{ \mathbf{h} \times 
\big[ \psi(\hat{D}\psi)^* + \psi^*(\hat{D}\psi) \big] \right\},
		\end{aligned}
		\label{eq:BuzdinF}
	\end{equation}

\noindent where \(\psi\) is the superconducting order parameter, \(a = \alpha(T - T_{c0})\), \(\gamma > 0\), and \(b > 0\) are the standard GL coefficients, and \(\hat{D}_i = -i\partial_i - 2eA_i\). The unit vector \(\mathbf{n}\) denotes the direction of the Rashba field (the gradient of the asymmetric potential), and \(\mathbf{h}\) is the exchange field in the ferromagnet. The coefficient \(\epsilon \propto \alpha_R\) characterizes the strength of the Rashba spin-orbit coupling. The final term, the  linear gradient coupling, breaks both inversion and time reversal symmetries.
	
	For nonreciprocal transport along the \(x\) direction, we use the geometry of Ref.~\cite{Buzdin2008}. In this configuration, n is along the $\hat{z}$ axis and the internal exchange field lies in the plane perpendicular to the current direction, \(\mathbf{h} = h\hat{y}\). The superconducting phase varies along \(x\). We assume that the orbital effect is negligible, so we set \(\mathbf{A} = 0\), which is a good approximation for thin films with in plane magnetization \cite{Buzdin2005}.
	
	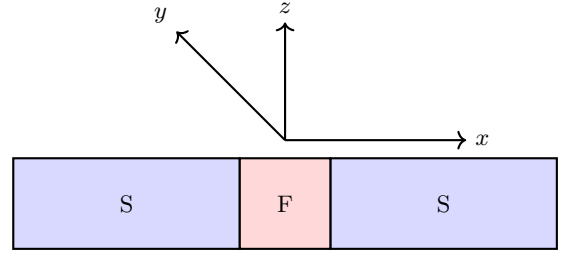
\begin{figure}[h]
		\centering
		\begin{tikzpicture}[scale=1.2]
			\filldraw[fill=blue!15, draw=black, thick] (-3,-0.5) rectangle (-0.5,0.5);
			\filldraw[fill=blue!15, draw=black, thick] (0.5,-0.5) rectangle (3,0.5);
			\filldraw[fill=red!15, draw=black, thick] (-0.5,-0.5) rectangle (0.5,0.5);
			\node at (-1.75,0) {S};
			\node at (1.75,0) {S};
			\node at (0,0) {F};
			\draw[->, thick] (0,0.7) -- (2,0.7) node[right] {$x$};
			\draw[->, thick] (0,0.7) -- (0,2) node[above] {$z$};
			\draw[->, thick] (0,0.7) -- (-1.2,1.9) node[above left] {$y$};
		\end{tikzpicture}
		\caption{Schematic picture of the SFS Josephson junction. The current flows along \(x\). The Rashba spin-orbit field is along \(\mathbf{n} = \hat{z}\), and the internal exchange field points along \(\mathbf{h} = h\hat{y}\).}
		\label{fig:geometry}
	\end{figure}

	The $\varphi_0$ junction model \cite{Buzdin2008} was formulated under the assumption of a weak exchange field $h \lesssim T_c$. This choice makes it possible to work within the standard gradient expansion and to obtain the current-phase relation $I = I_c \sin(\varphi - \varphi_0)$ without including higher order gradient terms. This approach works well for that parameter range\cite{Buzdin2008}.

In this strong field regime, the GL coefficient $\gamma$ can change sign and become negative over a finite range of parameters. When $\gamma<0$, the term $\gamma|\nabla\psi|^2$ lowers the free energy. The system therefore tends to develop a spatially modulated (oscillatory) state, which is precisely the origin of the damped oscillatory behavior of the Cooper pair wavefunction in a ferromagnet \cite{Buzdin2005}. In this case, without any higher order gradient contributions, the free energy functional becomes unbounded from below and the gradient energy can be made arbitrarily negative by increasing $|\nabla\psi|$, which is a manifestation of the fundamental instability of the gradient expansion.

To restore the stability of the functional and render it bounded from below, one must include a higher order derivative term with a positive coefficient $\eta>0$. This is the standard procedure for describing the proximity effect in the strong exchange field regime, as originally introduced in the generalized Ginzburg-Landau theory for superconductor-ferromagnet systems,
	\begin{equation}
		\begin{aligned}
			F ={}& a|\psi|^2 + \gamma|\hat{D}\psi|^2 + \frac{\eta}{2}|D^2\psi|^2 + \frac{b}{2}|\psi|^4 \\
			& - \epsilon\,\mathbf{n}\!\cdot\!\left\{ \mathbf{h} \times 
			\big[ \psi(\hat{D}\psi)^* + \psi^*(\hat{D}\psi) \big] \right\},
		\end{aligned}
		\label{eq:extended_Buzdin_F}
	\end{equation}
	The term $\frac{\eta}{2}|\hat{D}^2\psi|^2$ ensures that the free energy is bounded from below and sets the characteristic length scale for the spatial oscillations of the order parameter.

\section{asymmetric critical current}

With the order parameter written as $\psi(x) = |\psi| e^{i\phi(x)}$, the kinetic term in Eq.~\eqref{eq:extended_Buzdin_F} becomes $|\hat{D}_x\psi|^2 = q^2|\psi|^2$, where $q \equiv \partial_x\phi$. For the linear gradient coupling, we obtain
\begin{equation}
	\psi(\hat{D}_x\psi)^* + \psi^*(\hat{D}_x\psi) = 2q|\psi|^2.
\end{equation}

\begin{figure}[htbp]
	\centering
	\includegraphics[width=1.1\columnwidth]{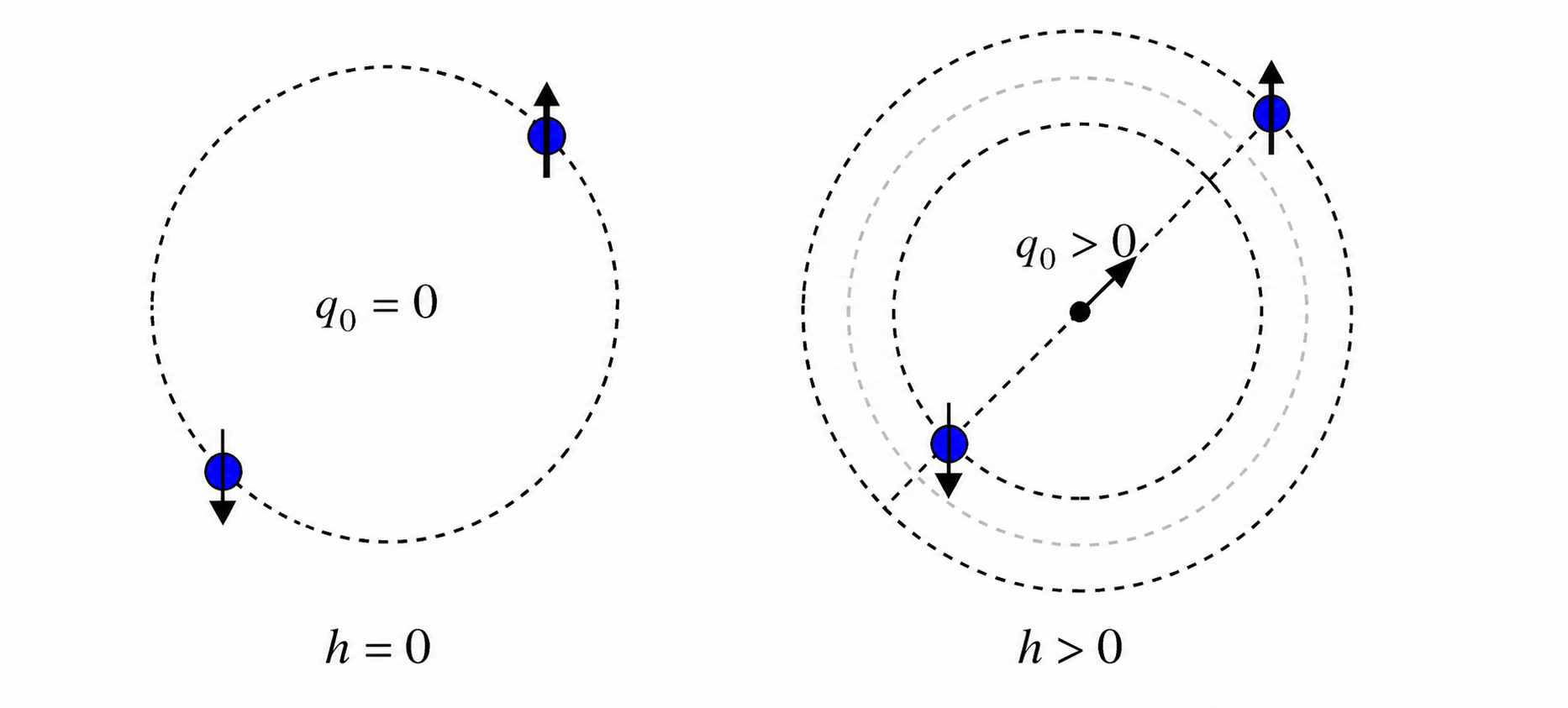}
	\caption{Picture of a Cooper pair in momentum space. Left: without exchange field, the electrons have opposite momenta, forming a zero momentum pair. Right: with an exchange field $h$, the spin polarization gives the pair a finite momentum $q_0$.}
	\label{fig:cooper_pair}
\end{figure}

The linear gradient coupling ($\propto \epsilon h q$) is responsible for finite momentum pairing. Without the higher order gradient term ($\eta = 0$), the free energy density is
\begin{equation}
	F(q) = (a + \gamma q^2 + 2\epsilon h q)|\psi|^2 + \frac{b}{2}|\psi|^4.
\end{equation}
Minimization with respect to $|\psi|^2$ yields $|\psi|^2 = -(a + \gamma q^2 + 2\epsilon h q)/b$. Minimization with respect to $q$ gives a nonzero ground state momentum $q_0 \approx -\epsilon h/\gamma$ for small $\epsilon h$. Thus the Lifshitz term alone produces a finite momentum state. Figure~\ref{fig:cooper_pair} illustrates how the exchange field alters the Cooper pair state in momentum space.

It is important to note that in \cite{Buzdin2008} the description of the $\varphi_0$ junction relies on two simplifying assumptions: a weak exchange field \(h \lesssim T_c\) and the neglect of the nonlinear term in the Ginzburg-Landau equation. Under these approximations, the current-phase relation reduces to the simple sinusoidal form \(I(\varphi) = I_c \sin(\varphi + \varphi_0)\), which is an exact solution for the current in that limit. In the present work, we retain the full nonlinear structure of the Ginzburg-Landau free energy without invoking these simplifications, and we determine the critical current by finding the extrema of the resulting current relation. This approach is similar to the analyses of Edelstein \cite{edelstein1996} and He \textit{et. al.} \cite{He2022}.

However, the current relation derived from this free energy remains symmetric: $I(q) = -I(-q)$, so $|I_{c+}| = |I_{c-}|$. The linear gradient coupling shifts the current along the $\phi$ axis (a $\phi_0$ shift) but does not break the symmetry $I(q) = -I(-q)$. To obtain a difference $\Delta I_c = |I_{c+}| - |I_{c-}|$, a higher order gradient term (e.g., a $q^4$ term) is necessary.

Substituting $\mathbf{n} = \hat{z}$ and $\mathbf{h} = h\hat{y}$ into Eq.~\eqref{eq:extended_Buzdin_F} yields
\begin{equation}
	\label{eq:Fq}
	F(q) = \left( a + \gamma q^2 + \frac{\eta}{2}q^4 + 2\epsilon h q \right)|\psi|^2 + \frac{b}{2}|\psi|^4.
\end{equation}

For a given phase gradient $q$, the amplitude $|\psi|^2$ is determined by minimizing the free energy with respect to $|\psi|^2$:
\begin{equation}
	\label{eq:amplitude}
	|\psi|^2 = -\frac{1}{b}\left( a + \gamma q^2 + \frac{\eta}{2}q^4 + 2\epsilon h q \right).
\end{equation}
This expression is valid provided the right hand side is positive, which restricts $q$ to the range where the system remains superconducting.

The supercurrent is obtained from the relation $I(q) = -2e\partial F/\partial q$. Using Eq.~\eqref{eq:amplitude} and noting that the contribution from $\partial|\psi|^2/\partial q$ vanishes by the minimization condition, we obtain
\begin{equation}
	\label{eq:Iq}
	I(q) = \frac{2e}{b} \left( 2\gamma q + 2\eta q^3 + 2\epsilon h \right) \
	\left( a + \gamma q^2 + \frac{\eta}{2}q^4 + 2\epsilon h q \right).
\end{equation}

To reveal the essential physics, we introduce dimensionless variables. Since $\gamma$ can in principle be negative in the strong exchange field regime, we use absolute values to ensure the definitions remain meaningful:
\begin{equation}
	\tilde{q} = q \sqrt{\frac{|\gamma|}{|a|}}, \qquad
	\tilde{\eta} = \frac{\eta |a|}{2\gamma^2}, \qquad
	\tilde{\kappa} = \frac{2\epsilon h}{\sqrt{|a||\gamma|}},
\end{equation}
and set $\tilde{I} = I / I_0$ with $I_0 = (2e/b)|a|\sqrt{|a|/|\gamma|}$. Below $T_c$, where $a = -|a|$ and assuming $\gamma>0$ for the moment (the case $\gamma<0$ is discussed below), Eq.~\eqref{eq:Iq} reduces to
\begin{equation}
	\label{eq:Itilde}
	\tilde{I}(\tilde{q}) = \left( 2\tilde{q} + 4\tilde{\eta}\tilde{q}^3 + \tilde{\kappa} \right)
	\left( 1 - \tilde{q}^2 - \tilde{\eta}\tilde{q}^4 - \tilde{\kappa}\tilde{q} \right).
\end{equation}

Equation~\eqref{eq:Itilde} is the central result of this section. It describes the current phase relation in dimensionless form. The critical currents $I_{c\pm}$ correspond to the extrema of $\tilde{I}(\tilde{q})$, which are obtained by solving $d\tilde{I}/d\tilde{q}=0$. As we demonstrate below, the asymmetry between $|I_{c+}|$ and $|I_{c-}|$ originates from the interplay between the linear Lifshitz term ($\tilde{\kappa}$) and the higher order gradient term ($\tilde{\eta}$).

\begin{figure*}[htbp]
	\centering
	\begin{subfigure}{0.48\textwidth}
		\centering
		\includegraphics[width=\textwidth]{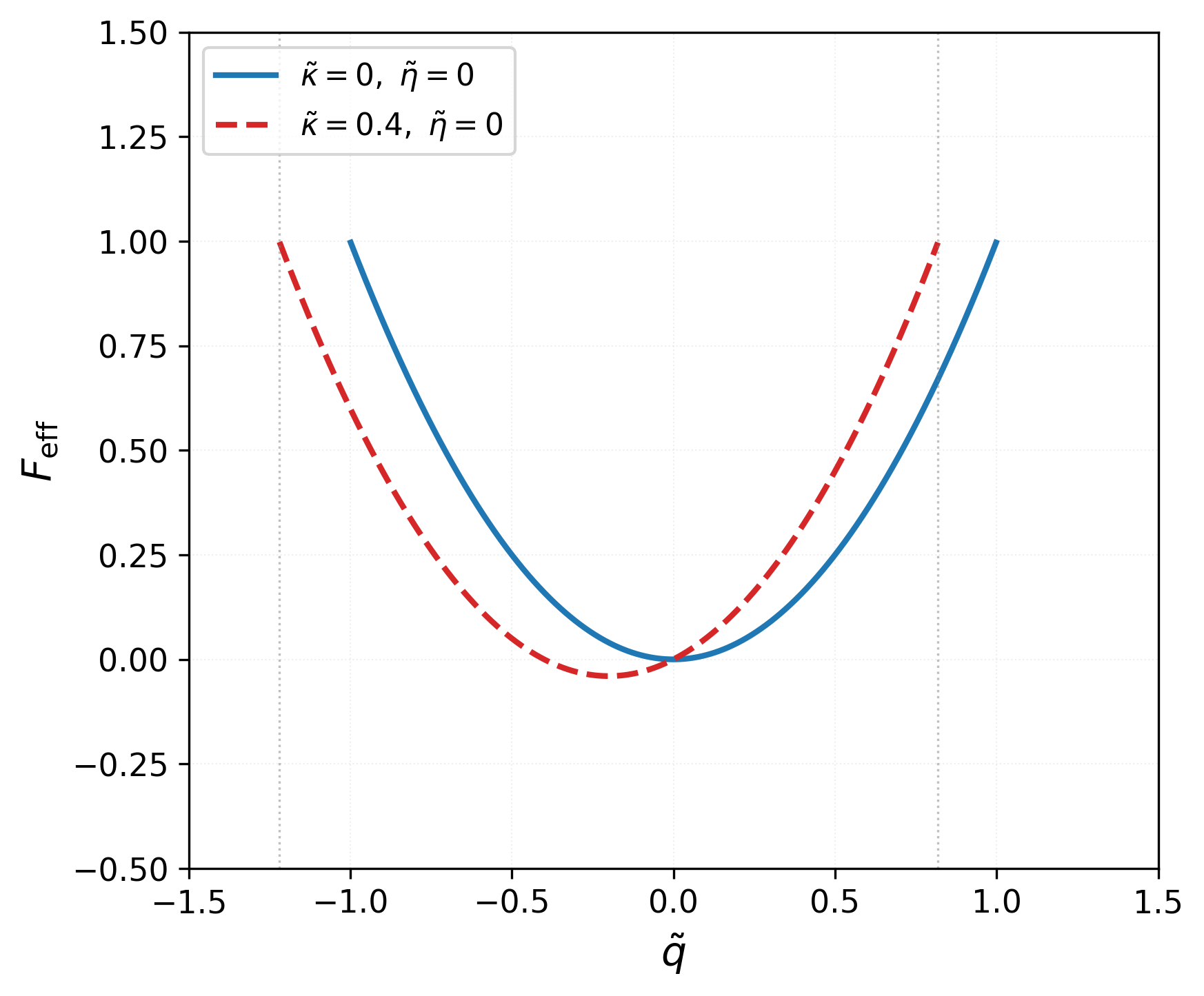}
		\caption{}
		\label{fig:free_weak}
	\end{subfigure}
	\hfill
	\begin{subfigure}{0.48\textwidth}
		\centering
		\includegraphics[width=\textwidth]{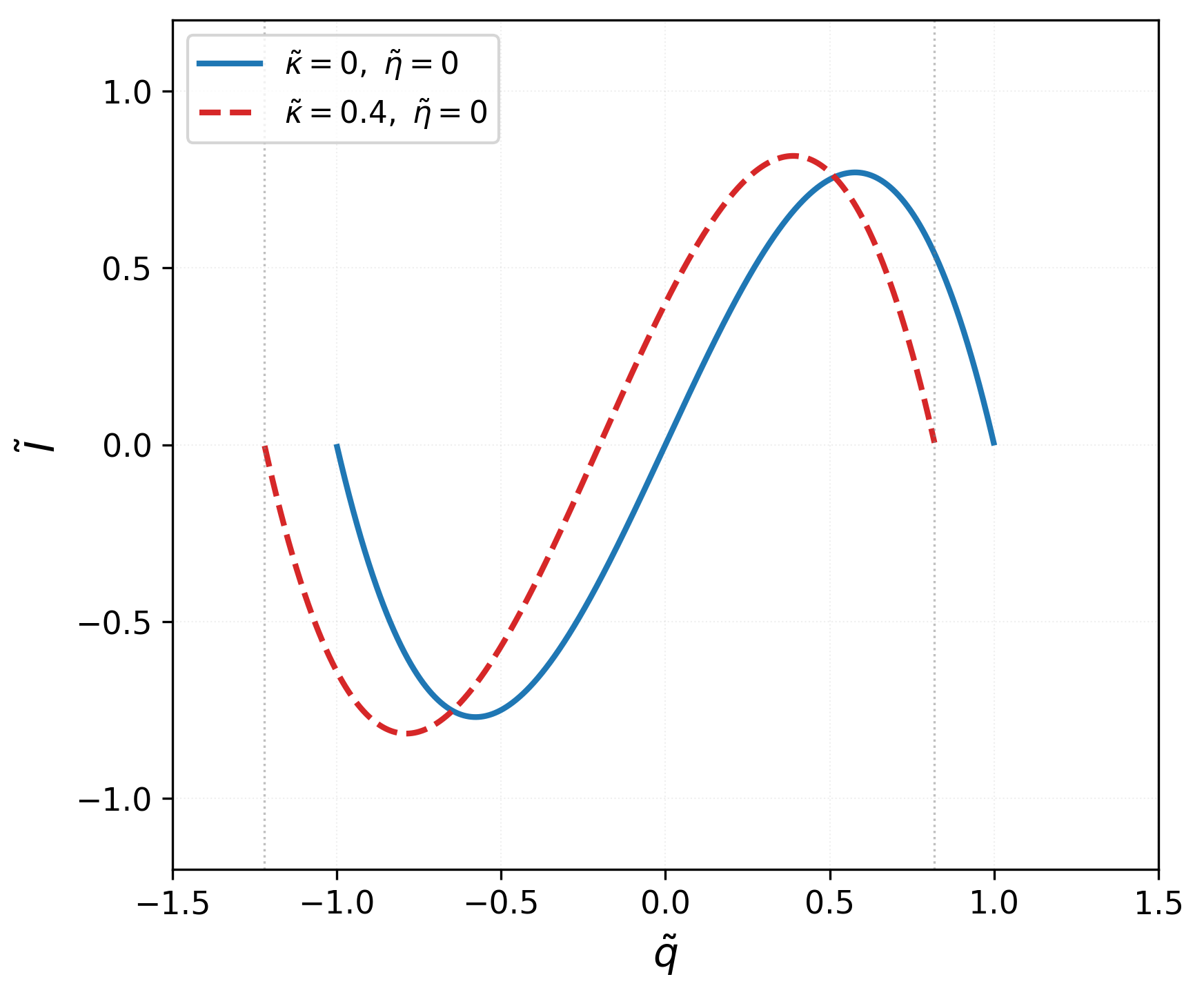}
		\caption{}
		\label{fig:current_weak}
	\end{subfigure}

	\par\medskip
	
	\begin{subfigure}{0.48\textwidth}
		\centering
		\includegraphics[width=\textwidth]{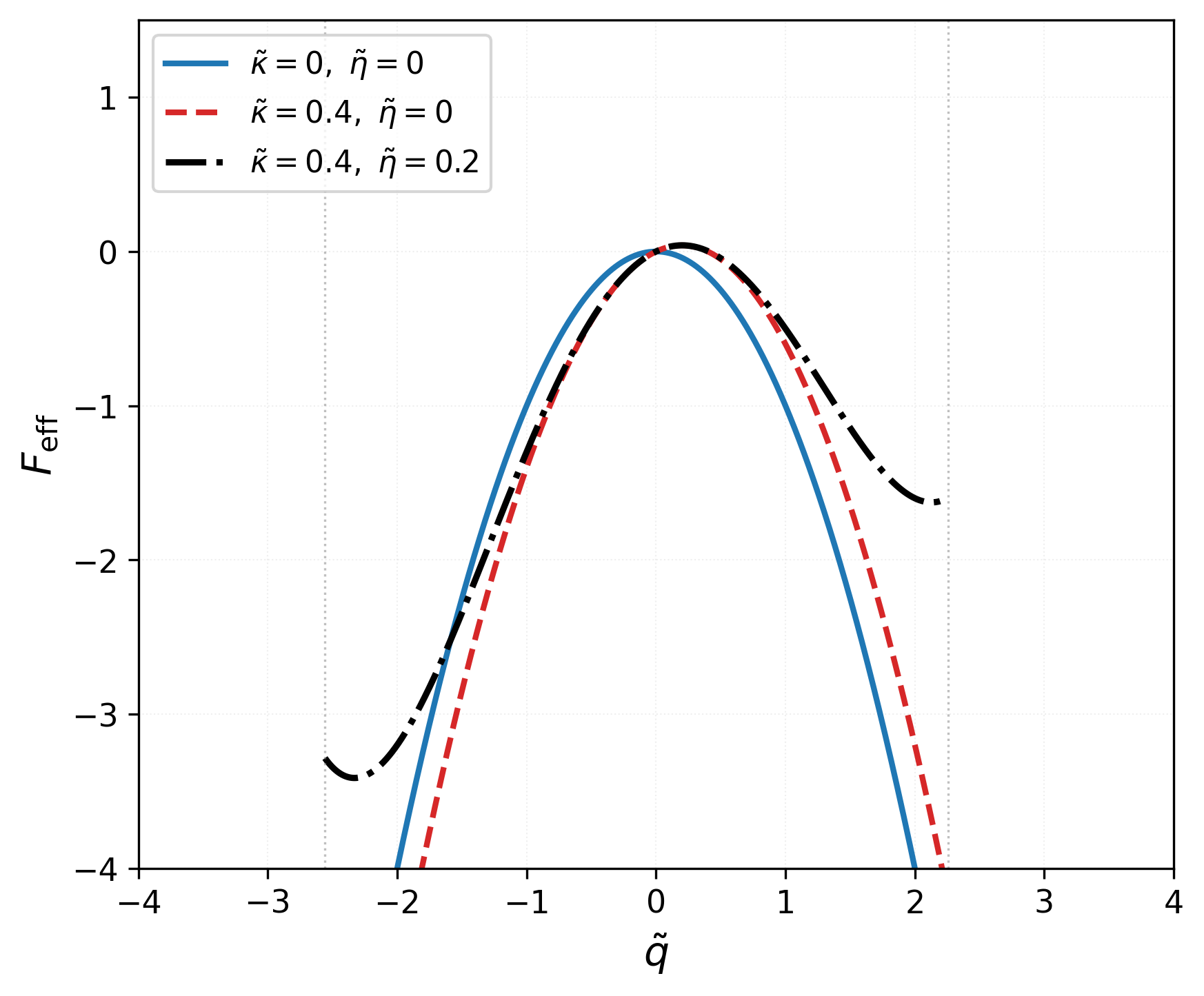}
		\caption{}
		\label{fig:free_strong}
	\end{subfigure}
	\hfill
	\begin{subfigure}{0.48\textwidth}
		\centering
		\includegraphics[width=\textwidth]{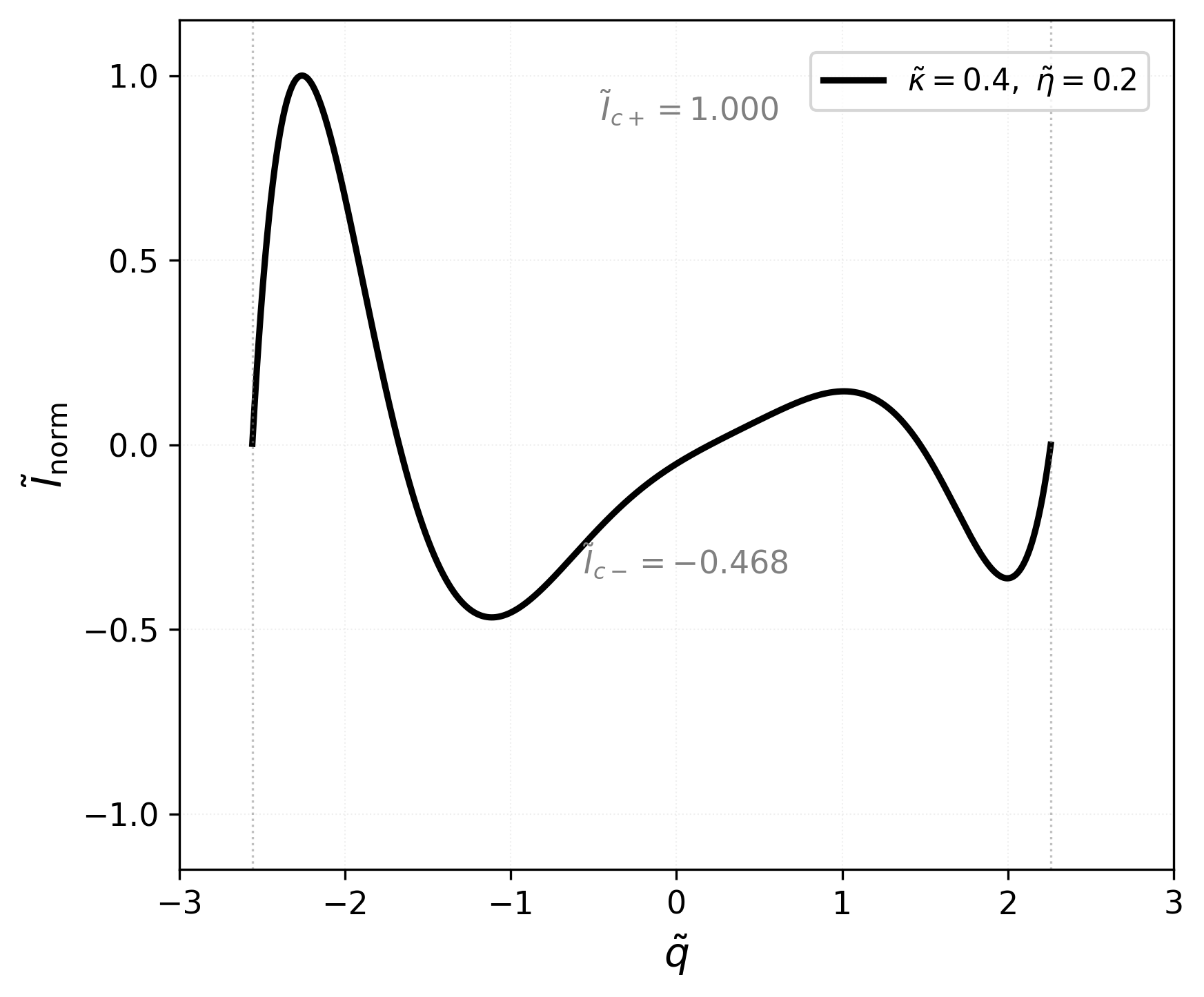}
		\caption{}
		\label{fig:current_strong}
	\end{subfigure}
	
	\caption{Free energy and current in the weak and strong field regimes. (a),(b) Weak field: $\gamma>0$, $\tilde{\eta}=0$, $\tilde{\kappa}=0$ and $0.4$. (c),(d) Strong field: $\gamma<0$, $\tilde{\kappa}=0.4$, $\tilde{\eta}=0.2$ (black dash dotted) and $\tilde{\eta}=0$ (red dashed). The current in (d) is normalized according to Eq.~\eqref{eq:Inorm}.}
\label{fig:four_panel}
\end{figure*}

Before analyzing the current, it is crucial to determine the physically allowed range of $\tilde{q}$. The superconducting state exists only where the amplitude squared is positive, $|\psi|^2 > 0$. In dimensionless form, Eq.~\eqref{eq:amplitude} gives the condition
	\begin{equation}
		\label{eq:q_range_weak}
		1 - \tilde{q}^2 - \tilde{\eta}\tilde{q}^4 - \tilde{\kappa}\tilde{q} > 0.
	\end{equation}
	For the weak field regime with $\tilde{\eta}=0$, this reduces to $1 - \tilde{q}^2 - \tilde{\kappa}\tilde{q} > 0$, which is a quadratic inequality. The roots are $\tilde{q} = (-\tilde{\kappa} \pm \sqrt{\tilde{\kappa}^2 + 4})/2$. For the parameter $\tilde{\kappa}=0.4$ used in Fig.~\ref{fig:four_panel}(a,b), the allowed interval is approximately $-1.22 < \tilde{q} < 0.82$. Outside this range, $|\psi|^2$ becomes negative, indicating that the superconducting state has disappeared. Since the current $I$ is proportional to $|\psi|^2$, it vanishes at the boundaries of this interval.

For the strong field regime, where $\gamma<0$, the dimensionless current takes the form
	\begin{equation}
		\label{eq:Itilde_strong}
		\tilde{I}(\tilde{q}) =
		\left( -2\tilde{q} + 4\tilde{\eta}\tilde{q}^3 + \tilde{\kappa} \right)
		\left( -1 - \tilde{q}^2 + \tilde{\eta}\tilde{q}^4 + \tilde{\kappa}\tilde{q} \right).
	\end{equation}
	The condition $|\psi|^2 > 0$ now reads
	\begin{equation}
		\label{eq:q_range_strong}
		-1 - \tilde{q}^2 + \tilde{\eta}\tilde{q}^4 + \tilde{\kappa}\tilde{q} > 0.
	\end{equation}
	For $\tilde{\eta}=0.2$ and $\tilde{\kappa}=0.4$, as used in Fig.~\ref{fig:four_panel}(c,d), solving this quartic inequality numerically gives the allowed interval $-2.6 < \tilde{q} < 2.25$. Again, the current vanishes at these boundaries.

These allowed ranges have a clear physical meaning: they correspond to the region where the superconducting order parameter is nonvanishing. The boundaries $\tilde{q}_{\pm}$ are the points where $|\psi|^2 = 0$; i.e. the superconducting to normal transition induced by the finite momentum. By restricting our plots to these intervals, we ensure that only physically accessible states are shown and that the computed critical currents $I_{c\pm}$ lie within the superconducting region.

For the weak field regime shown in Fig.~\ref{fig:four_panel}(a) and Fig.~\ref{fig:four_panel}(b), we set $\tilde{\eta}=0$. The free energy in Fig.~\ref{fig:four_panel}(a) is obtained from Eq.~\eqref{eq:Fq}, while the current in Fig.~\ref{fig:four_panel}(b) is obtained from Eq.~\eqref{eq:Itilde}. The Lifshitz term shifts the minimum of the free energy to a finite momentum, while the symmetry of the current remains unchanged.

For the strong field regime, where $\gamma<0$, Eq.~\eqref{eq:Iq} takes the dimensionless form given in Eq.~\eqref{eq:Itilde_strong}. The free energy shown in Fig.~\ref{fig:four_panel}(c) is obtained from Eq.~\eqref{eq:Fq} with $\gamma<0$. In this case, the higher order gradient term is required to keep the free energy bounded from below. The current shown in Fig.~\ref{fig:four_panel}(d) is obtained from Eq.~\eqref{eq:Itilde_strong}. Since Eq.~\eqref{eq:amplitude} determines the region in which the superconducting state exists, only the range satisfying $|\psi|^2>0$ is used in Fig.~\ref{fig:four_panel}(d).

For graphical comparison, the current in Fig.~\ref{fig:four_panel}(d) is additionally normalized according to
\begin{equation}
	\label{eq:Inorm}
	\tilde{I}_{\mathrm{norm}} =
	\frac{\tilde{I}}
	{\max\left(|\tilde{I}_{c+}|,|\tilde{I}_{c-}|\right)}.
\end{equation}
This normalization only changes the scale used for plotting and does not change the relative difference between the positive and negative critical currents.

Figure~\ref{fig:four_panel} summarizes the difference between the two regimes. In the weak field regime, the Lifshitz term produces a finite momentum state without changing the symmetry of the critical currents. In the strong field regime, the higher order gradient term stabilizes the free energy and, together with the Lifshitz coupling, produces the asymmetric critical currents discussed below.

\section{Josephson Diode Effect}

The physical origin of the diode effect is illustrated in Fig.~\ref{fig:four_panel}.
Figures (a) and (b) correspond to the weak field regime ($h\lesssim T_c$), where
$\gamma>0$ and the higher order term is absent ($\eta=0$). In this case, the Lifshitz
invariant only tilts the free energy and shifts the current
along the $\phi$ axis, but the symmetry $I(q)=-I(-q)$ is preserved, resulting
in equal critical currents.

By contrast, in the strong field regime ($h\gg T_c$,
$\gamma<0$), the gradient expansion without the $\eta$ term is unstable, as shown
by the red dashed curve in Fig.3(c). Adding the quartic term with $\eta>0$
(black dash dotted curve) stabilizes the free energy and creates an asymmetric
potential well. Consequently, the current in figure (d) exhibits
unequal critical currents $I_{c+} \neq |I_{c-}|$,
directly demonstrating the diode effect. Thus, the interplay between the
linear gradient coupling and the higher order term is essential for the SDE.

As shown in Fig.~\ref{fig:four_panel}(a), the weak field free energy is obtained from Eq.~\eqref{eq:Fq} with $\gamma>0$ and $\eta=0$. The corresponding current in Fig.~\ref{fig:four_panel}(b) is obtained from Eq.~\eqref{eq:Itilde}. The two curves in Fig.3(a) show that the Lifshitz term shifts the free energy minimum, while Fig.3(b) shows that the corresponding critical currents remain symmetric.

In the strong field regime, the free energy in Fig.~\ref{fig:four_panel}(c) is obtained from Eq.~\eqref{eq:Fq} with $\gamma<0$ and $\eta>0$. The red dashed curve corresponds to the case $\eta=0$ and becomes unbounded from below, while the black curve remains bounded due to the higher order gradient term. The resulting asymmetric free energy is reflected in the current shown in Fig.~\ref{fig:four_panel}(d).

The current in Fig.~\ref{fig:four_panel}(d) is calculated from Eq.~\eqref{eq:Itilde_strong} only in the superconducting region determined by Eq.~\eqref{eq:amplitude}. For graphical comparison, the current is additionally normalized according to Eq.~\eqref{eq:Inorm}. The normalization changes only the vertical scale of the plotted current and does not affect the inequality between the two critical currents. The resulting difference $I_{c+}\neq |I_{c-}|$ demonstrates the nonreciprocal response.

The diode quality factor is defined as
\begin{equation}
	\label{eq:Q_def}
	Q \equiv \frac{|I_{c+}| - |I_{c-}|}{|I_{c+}| + |I_{c-}|}.
\end{equation}

For small $\tilde{\kappa}$ and $\tilde{\eta}$, a perturbative expansion about the extrema of the symmetric case ($\tilde{\kappa} = \tilde{\eta} = 0$) yields
\begin{equation}
	\label{eq:Ic+}
	\tilde{I}_{c\pm} = \pm \frac{4}{3\sqrt{3}} \left( 1 \mp \frac{5}{2\sqrt{3}} \tilde{\kappa} \tilde{\eta} \right).
\end{equation}

A nonzero difference $\Delta I_c =| I_{c+}|-|I_{c-}|$ already guarantees
diode behaviour: for a bias current chosen between $I_{c-}$ and
$I_{c+}$ the junction is resistive for one direction and
superconducting for the opposite.

Substituting  Eq.~\eqref{eq:Ic+} this into Eq.~\eqref{eq:Q_def} gives
\begin{equation}
	\label{eq:Q}
	Q = \frac{5}{2\sqrt{3}}  |\tilde{\kappa} \tilde{\eta}| =
	\frac{5}{2\sqrt{3}}  \epsilon \eta h  \frac{|a|^{1/2}}{|\gamma|^{5/2}}.
\end{equation}

\begin{figure}[htbp]
	\centering
	\includegraphics[width=\columnwidth]{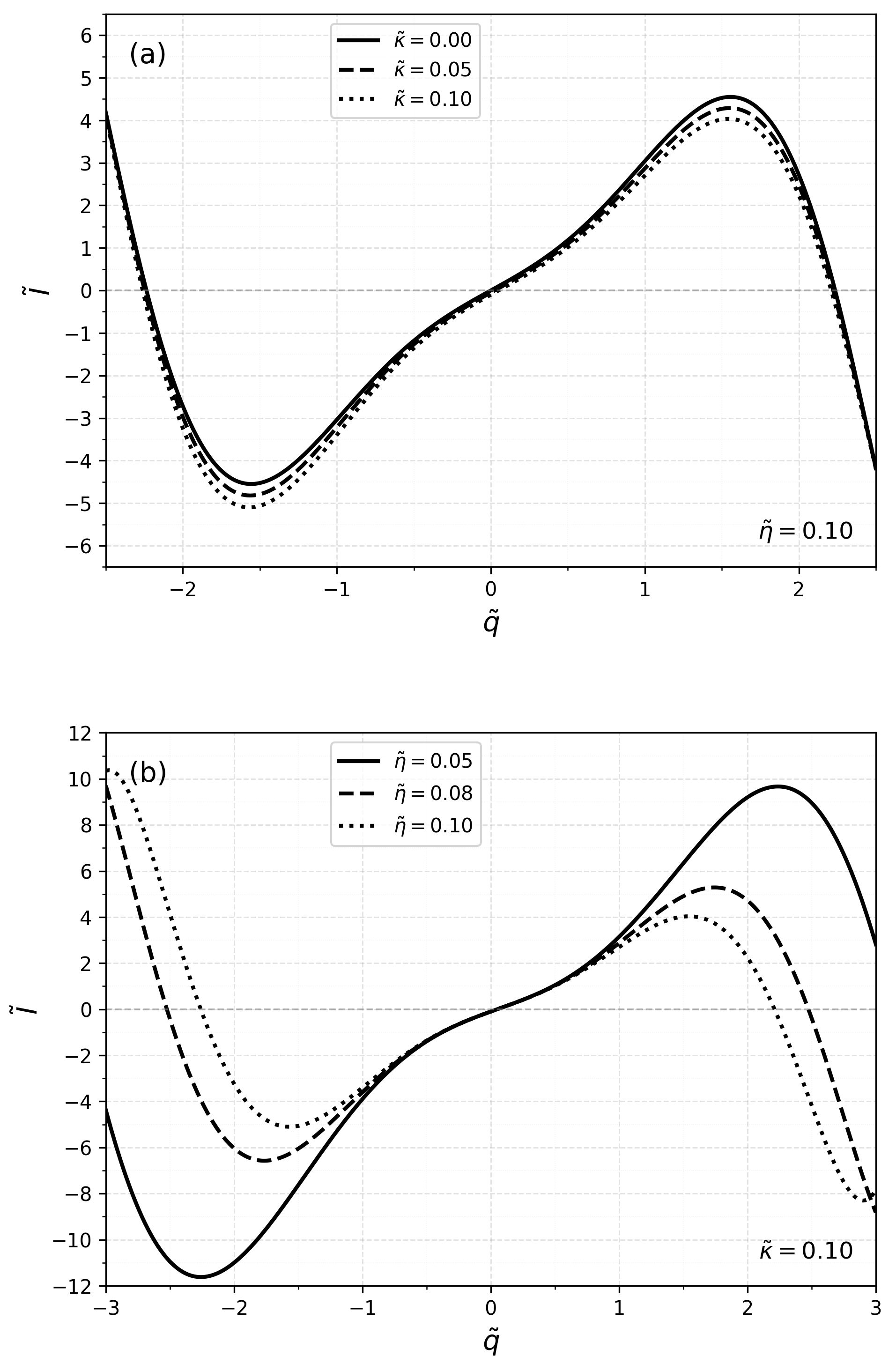}
	\caption{Dimensionless current $\tilde{I}(\tilde{q})$ at the (a) different $\kappa$ and (b) different $\eta$, in the strong field regime, plotted only in the superconducting interval $|\psi|^2>0$.}
	\label{fig:current_kappa_eta}
\end{figure}

The absolute value ensures that $Q$ is positive regardless of the sign of $\gamma$. Importantly, Eq.~\eqref{eq:Q} shows that the diode effect vanishes when either the Lifshitz coefficient $\epsilon$ or the higher order coefficient $\eta$ is zero.
To see how the nonreciprocity builds up with the two parameters, we show in Fig.~\ref{fig:current_kappa_eta} the dimensionless current $\tilde{I}(\tilde{q})$ in the strong field regime for several values of $\tilde{\kappa}$ and $\tilde{\eta}$. In Fig.4(a), the higher order coefficient is fixed at $\tilde{\eta}=0.10$, and the Lifshitz parameter is varied as $\tilde{\kappa}=0, 0.05, 0.10$. In Fig.4(b), the Lifshitz parameter is fixed at $\tilde{\kappa}=0.10$, and the higher order coefficient is varied as $\tilde{\eta}=0.05, 0.08, 0.10$. All curves are shown only within the superconducting interval where $|\psi|^2>0$. When $\tilde{\kappa}=0$ (solid curve in Fig.4a), the current is odd under $\tilde{q}\to-\tilde{q}$ and the two critical currents are equal. As $\tilde{\kappa}$ increases, the positive and negative extrema become increasingly different, signaling the onset of the diode effect. Similarly, Fig.4(b) shows that increasing $\tilde{\eta}$ at fixed $\tilde{\kappa}$ amplifies the asymmetry. These trends are consistent with Eq.~\eqref{eq:Q}: the diode quality factor grows linearly with both $\tilde{\kappa}$ and $\tilde{\eta}$, and it vanishes when either of them is zero. This confirms that the coexistence of the linear gradient coupling and the higher order gradient term is the essential ingredient for the superconducting diode effect in the strong exchange field regime.
To understand the asymmetry it is helpful to look at the free energy landscape.

	\section{Conclusion}
We have shown that the superconducting diode effect in a $\varphi_0$ Josephson junction can be extended to the strong exchange field regime by including a higher order gradient term in the Ginzburg Landau free energy. While the linear gradient coupling generates finite momentum pairing, it does not by itself produce nonreciprocal critical currents. The diode effect arises from the interplay between the linear gradient coupling and the higher order gradient term, leading to $I_{c+}\neq |I_{c-}|$ and a finite diode quality factor. 
	\section{Acknowledgements}
	 A. Janalizadeh and M. R. Kolahchi acknowledge support from IASBS. Yu. M. Shukrinov acknowledges the support by the Russian Science
	Foundation (Grant 22-71-10022).

\end{document}